\documentclass{emulateapj}
\usepackage{natbib}
\usepackage{amsmath}
\usepackage{graphics,epsfig}
\usepackage{url}
\usepackage[usenames,dvipsnames]{xcolor}
\usepackage{multirow}

\newcommand{\degree}{\ensuremath{^\circ}}
\newcommand{\superscript}[1]{\ensuremath{^{\textrm{#1}}}}
\newcommand{\subscript}[1]{\ensuremath{_{\textrm{#1}}}}
\renewcommand{\i}[1]{\textit{#1}}
\def\spose#1{\hbox to 0pt{#1\hss}}
\def\lta{\mathrel{\spose{\lower 3pt\hbox{$\mathchar"218$}}
     \raise 2.0pt\hbox{$\mathchar"13C$}}}
\def\gta{\mathrel{\spose{\lower 3pt\hbox{$\mathchar"218$}}
     \raise 2.0pt\hbox{$\mathchar"13E$}}}

\shorttitle{Radio Detection of Multiple Exomooons}
\shortauthors{Noyola Satyal Musielak}

\begin{document}

\title{On the radio detection of multiple-exomoon systems due to plasma torus sharing}

\author{J. P. Noyola, S. Satyal and Z.E. Musielak}
\affil{The Department Physics, University of Texas at Arlington, Arlington, TX 76019;\\ 
       jpnoyola@uta.edu; ssatyal@uta.edu; zmusielak@uta.edu}

\begin{abstract}
The idea of single exomoon detection due to the radio emissions caused by its interaction with the host exoplanet is extended to multiple-exomoon systems.
The characteristic radio emissions are made possible in part by plasma from the exomoons own ionosphere (Noyola et. al. 2014). In this work, it is demonstrated that neighboring exomoons and the exoplanetary magnetosphere could also provide enough plasma to generate a detectable signal. In particular, the plasma-torus-sharing phenomenon is found to be particularly well suited to facilitate the radio detection of plasma-deficient exomoons. The efficiency of this process is evaluated, and the predicted power and frequency of the resulting radio signals are presented.
\end{abstract} 


\section{Introduction}\label{sec:intro} 
Exomoons continue to remain elusive despite the steady search for over a decade. The previous suggestion by 
\cite{kip09} that exomoons may actually be discovered in the data from the Kepler mission still remains unconfirmed.
Moreover, the exomoon candidates reported by \cite{benj14}, and \cite{ben14} have not been independently verified.
Thus, the search for exomoons continues.
In our previous work \citep{noy14}, we explained how the interactions between an exomoon, a plasma torus, and the magnetosphere of Jovian exoplanet can lead to detectable radio emissions, which reveal the presence of the exomoon. 
In the current paper, we show that these type of interactions can be induced by plasma sources other than the exomoon’s own ionosphere, particularly another exomoon's plasma torus, and that using the plasma from those sources can make 
an exomoon radio detectable.
We explain how the plasma torus sharing mechanism can readily lead to multiple exomoon detections if (i) the plasma source is also a radio detectable exomoon, or (ii) if two or more exomoons share the same plasma and produce enough power.
Therein lies the importance of sharing plasma: it makes it possible to find 2 or more exomoons in a system where we would otherwise only find one.

This paper is organized as follows.
In Sec. \ref{sec:basic}, we describe the basic mechanism of the Io-Jupiter system, detection through plasma torus sharing is discussed in Sec. \ref{sec:share}, the frequency and temporal characteristics of the expected radio signal are presented in Sec. \ref{sec:fC}, and the consequences of these characteristics are outlined in Sec. \ref{sec:gifts}.
Alternative plasma sources are discussed in Sec. \ref{sec:misc}, and our conclusions are presented in Sec. \ref{sec:bye}.

\section{Basic Mechanism}\label{sec:basic}

The idea that exomoons can be discovered with radio telescopes \citep{noy14} came from the observed interaction between Jupiter and its moon Io, which produces radio emissions in the tens of MHz \citep{hes08}.
Io's ionosphere, which is produced by volcanic activity \citep{lop07}, injects ions into Jupiter's magnetosphere.
The ions then co-rotate with the Jovian magnetic field in the plane of the magnetic equator, creating a plasma torus \citep{su09}.
The speed difference between Io and the co-rotating plasma creates a unipolar inductor \citep{gol69} that induces a current across Io's atmosphere of a few million Amperes.
Simultaneously, the interaction between Io and the plasma torus produces magnetic field oscillations, known as Alfv\'en waves \citep{bel87}.
These oscillations generate electric fields parallel to the Jovian magnetic field lines, which then transport electrons toward Jupiter's magnetic poles \citep{neu80, cra97, sau99, su09}.

As the electrons travel through the field lines, they emit decametric radio emissions known as Io-DAM, through the electron cyclotron maser instability mechanism \citep{cra97, mau01}.
The plasma in Jupiter's ionosphere then transports the electrons between current-carrying field lines, thus completing a closed circuit with Io as the source of the electromotive force. 
Due to Io's role in the circuit, we will refer to it, and all moons and exomoons of this type, as \i{electromotive 
moons}.
The other Galilean moons, as well as Titan and Enceladus on Saturn, are also electromotive moons.
Thus, electromotive moons are common in our solar system, and may be common in other planetary systems as well.
It is noteworthy that while volcanism is essential to the formation of a dense Ionian ionosphere, such a process may not be required for larger moons that can sustain thick atmospheres (e.g. Titan), which in turn can create an ionosphere.

\begin{figure}
\epsscale{0.80}
\includegraphics[width=1\linewidth]{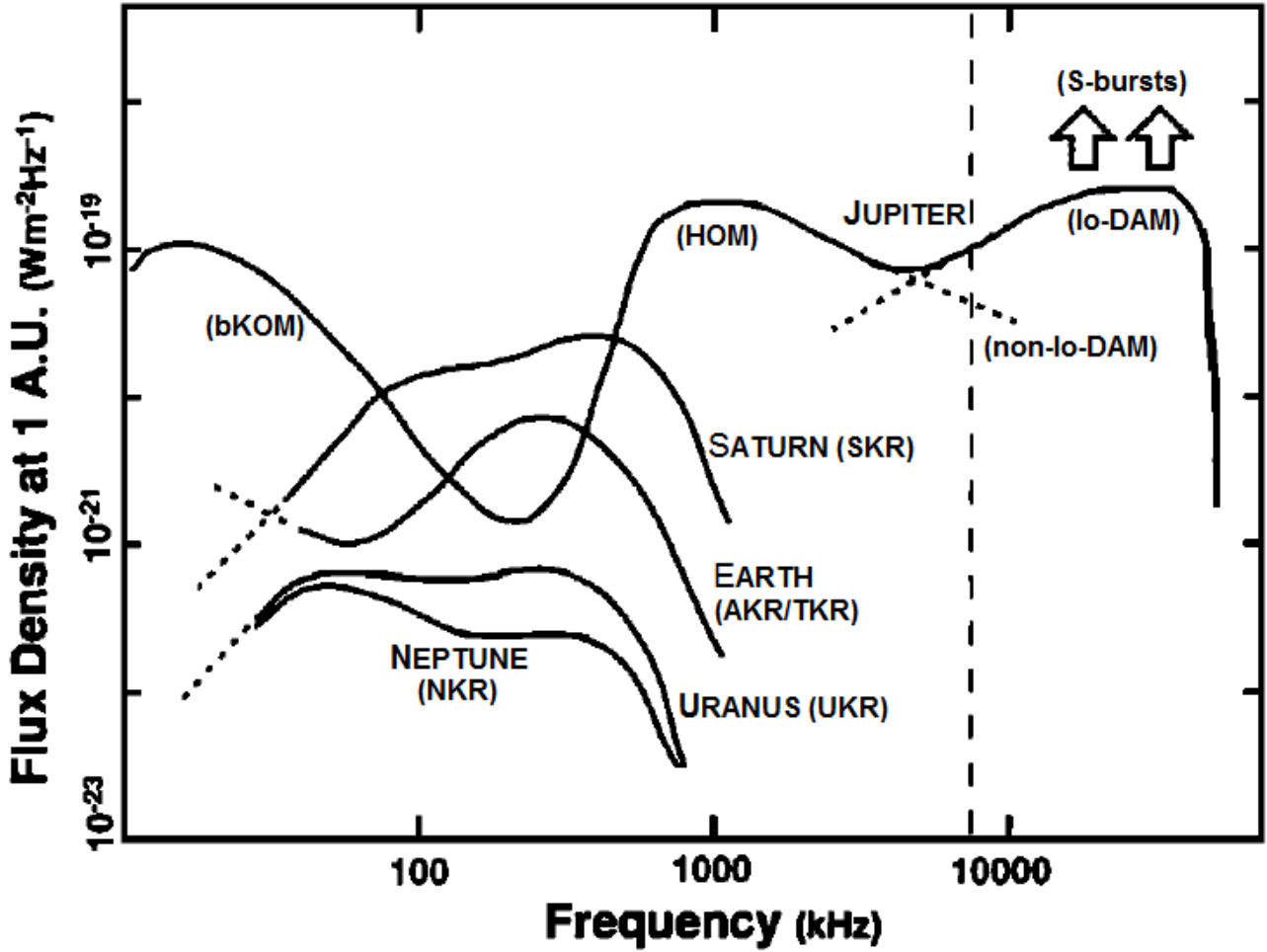}\hfill
\caption{Summary of planetary radio emissions found in the Solar System. Taken from \cite{zar98}}
\label{fig:radios}
\end{figure}

There are also other planetary radio emissions in the solar system that are not related to the presence of satellites, including the Jovian decametric emissions known as non-Io-DAM which occur at the same frequencies as Io-DAM emissions.
Figure \ref{fig:radios}, taken from \cite{zar98}, shows a summary of all planetary radio emissions found in our Solar System.
The Io-DAM, and non-Io-DAM emissions are shown to the right of the ionospheric cutoff, and where it is clear that Io-DAM radio emissions are considerably stronger than non-Io-DAM emissions for most of the applicable frequencies.
Later it will also be shown that these two emission types have completely different dynamic spectra as well.
As for the lower frequencies, only those which are higher than the ionospheric cut-off are relevant to this study because those are the only frequencies that are detectable from the ground.

When extrapolating the dynamics of Io and Jupiter to exomoons, we used the power relationship found by \cite{neu80}, and reported in \cite{noy14} as

\begin{equation}
P_{S}=\frac{\pi\beta_{S}R_{S}^{2}B_{S}^{2}V_{\rho}}{\mu_{0}}\sqrt{\frac{\rho_{S}}
{\rho_{S}+\frac{1}{\mu_{0}}\left(\frac{B_{S}}{V_{\rho}}\right)^{2}}},
\label{eq:P_S}
\end{equation}

\noindent
where $\beta_{S}$ is an efficiency coefficient ($\approx 1\%$), $R_{S}$ is the exomoon radius, $B_{S}$ is the local magnetic field, $V_{\rho}$ is the plasma speed with respect to the exomoon, $\rho_{S}$ is the local plasma density, and $\mu_{0}$ is the permeability of free space.
The dependence of the emission power on the exomoon's cross-sectional area ($\pi R_{S}^{2}$) leads to the 
question of how large exomoons can be.  
Formation models suggest that moons can form around Jovian planets with masses up to $10^{-4}$ times the mass of their host planet \citep{can06,hel15}.
Therefore large exoplanets of $3 M_{J}$ or larger could form a Mars-size exomoon, and $10 M_{J}$ exoplanets could have several Mars-size exomoons or larger \citep{hel15} (where $1 M_{J}$ is one Jupiter mass).
In contrast, smaller Jovian exoplanets are unlikely to harbor large exomoons, unless those moons are captured \citep{hel15}.  Hence, large Jovian exoplanets are of primary importance to the existence of large exomoons, and consequently to the mechanism described here.

To calculate the magnetic field we assume that the field is mostly dipolar, as is approximately the case for all magnetized bodies in the solar system, and that its angle of inclination with respect to the axis of rotation (referred to as the z-axis hereafter) is small enough to be neglected.
In other words, we assume that the $x$ and $y$ components of the dipole moment are negligible, which is a reasonable assumption since, even at a 20{\degree} inclination, the $z$ component constitutes 94\% of the total magnitude.
If the exoplanet has a magnetic dipole moment $\vec{m}_{P}=m_{P}\hat{z}$, then the exoplanet's magnetic field (in spherical coordinates) is given by

\begin{equation}
B(\vec{r})=\frac{\mu_{0}}{4\pi}\frac{m_{P}}{r^{3}}\left(3\cos^{2}\theta+1\right)^{\frac{1}{2}}.
\label{eq:Bmag}
\end{equation}

The approximation for the magnetic dipole moment is model taken from \cite{dur09}, and expressed here as

\begin{equation}
m_{P}=m_{J}\left(\frac{M_{P}}{M_{J}}\frac{T_{J}}{T_{P}}\right)^{K},
\label{eq:m}
\end{equation}

\noindent
where $M_{P}$ and $T_{P}$ are the exoplanet's mass and rotation period, $M_{J}=1.8986\times10^{27}$ kg, $T_{J}$ is Jupiter's rotation period ($9.925$ hours), $m_{J}$ is Jupiter's magnetic dipole moment ($1.56\times10^{27}$ A•m\superscript{2}), and $K$ is an experimental constant set to 1.15 \citep{noy14} to best fit the data from Solar System's giant planets.

The model assumes that the exoplanetary magnetic field rotates at the same rate as the exoplanet, as is the case for Jupiter, and that the rotation period, $T_{P}$, is also the same as Jupiter's.
How to predict the spin rate of planets is still an open debate in planet formation theory, but currently the most accurate fit to the observational data is one of a linear relationship between $log(M_{P})$ and the logarithm of a planet's spin angular momentum \citep{hug03}.
In other words, it would not be unreasonable to assume that bigger $M_{P}$ means smaller $T_{P}$, especially after $\beta$ Pictoris b's rotation period was found to be $\approx8.1\pm1.0$ hours \citep{sne14}, or $\approx18\%$ faster than Jupiter. 
Additionally, we can see from Equation \ref{eq:m} that a smaller period leads to a larger magnetic moment, which in turn leads to a larger $B_{S}$, and thus a higher emission power $P_{S}$.
Therefore, assuming that $T_{P}=T_{J}$ is a conservative assumption, and should work well for our purposes.

The relative plasma velocity, $V_{\rho}$, has two components: the exomoon's orbital velocity, and the corotation velocity.
We assume the exomoon is in a near circular orbit of radius $r_{S}$, so its orbital velocity, $V_{orb}$, is given by,

\begin{equation}
V_{orb}=\sqrt{GM_{P}/r_{S}}.
\label{eq:vorb}
\end{equation}

\noindent
To calculate the corotation velocity, $V_{co}$, we use the fit to Jupiter's observational data

\begin{equation}
V_{co}=(12.6\mbox{ km/s})\left(1.12-\frac{r_{S}}{50R_{J}}\right)\left(\frac{r_{S}}{R_{J}}\right),
\label{eq:vco}
\end{equation}

\noindent
found by \cite{bag11}, and which is valid for distances up to $\approx 28 R_{J}$.
The plasma corotates with the planetary magnetic field in near-rigid corotation (which in turn rotates with exoplanet), so the plasma's rotation is always prograde, unlike the exomoon which can be either prograde or retrograde.
Exomoon plasma tori have not been well studied.
To the best of our knowledge, as of the time of this study \cite{joh06} and \cite{benj14} are the only articles fully dedicated to exomoon plasma tori.
Hence, we will not attempt to generalize Equation \ref{eq:vco} for other exomoon plasma tori at this time.

\begin{figure}
\epsscale{0.80}
\includegraphics[width=1\linewidth]{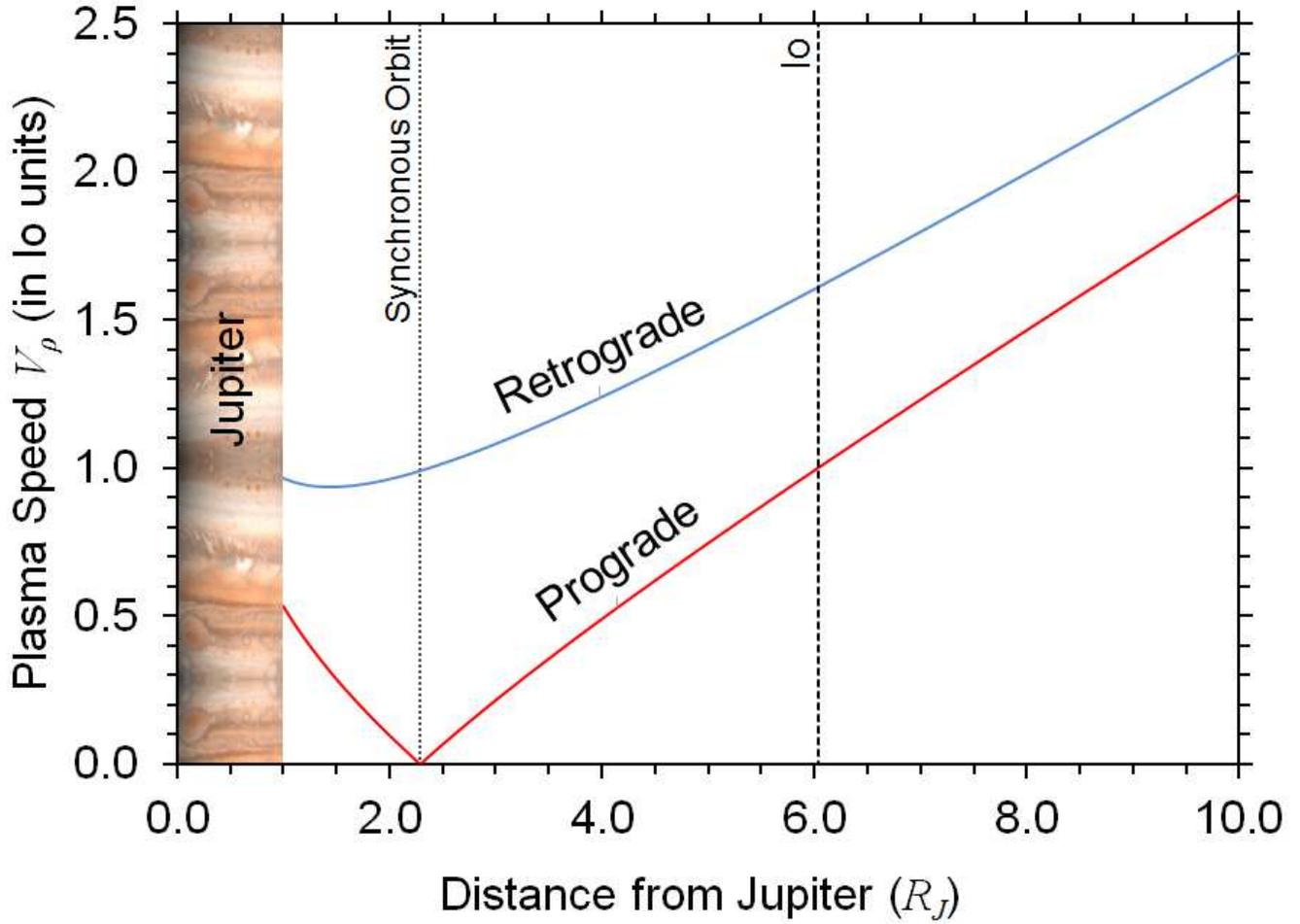}\hfill
\caption{The relative speed between the co-rotating plasma and a Jovian moon, $V_{\rho}$, for orbits between 1 and 10 $R_{J}$, expressed in Io units (1 unit $\approx$ 56.8 km/s).
The node seen for the prograde case at the synchronous orbit ($\approx 2.29$ $ R_{J}$) means that prograde exomoons in near-synchronous orbits cannot be detected. Io's orbit and Jupiter's boundary are shown for reference.}
\label{fig:Vabs}
\end{figure}

Using Equation \ref{eq:vco}, we find that $V_{\rho}$ is given by

\begin{equation}
V_{\rho}=(12.6\mbox{ km/s})\left(1.12-\frac{r_{S}}{50R_{J}}\right)\frac{r_{S}}{R_{J}}\mp\sqrt{\frac{GM_{P}}{r_{S}}},
\label{eq:vrho}
\end{equation}

\noindent
where the $'-'$ and $'+'$ signs correspond to prograde and retrograde exomoon orbits, respectively.
The direction of rotation of the exomoon makes a big difference.
As shown in Equation \ref{eq:vrho} and Figure \ref{fig:Vabs}, retrograde exomoons  will have larger $V_{\rho}$ than prograde exomoons. 
Since $P_{S}$ is proportional to $V_{\rho}$, we can conclude that retrograde exomoons will generate more radio power, and thus have a better chance to be detected.
Additionally, $V_{\rho}=0$ when prograde exomoons are near the synchronous orbit (see Figure \ref{fig:Vabs}), so $P_{S}=0$ as well, making detections on these orbits impossible.
The retrograde case does not have this problem, although $V_{\rho}$ does have a minimum value close to the planet's surface, as we can see from the figure.

Since retrograde orbits have significant advantages over prograde orbits, a few things about their existance must be noted.
In the Solar System, the Neptunian moon Triton is the only large moon in a retrograde orbit, and it is precisely due to its orbit that it has long been suspected to be a captured moon \citep[ and references therein]{agn06}.
In fact, numerical studies suggest that captured moons tend to stay in retrograde orbits about 50\% of the time \citep{por11}.
Thus, capture events provide a clear path by which exomoons can end up in a retrograde orbit.
For a more comprehensive review of exomoon capture see \cite{hel14}.

Now that we have a way to calculate the emitted radio power, we can proceed to calculate the emission's flux density. If the exomoon is a distance $d$ from Earth, then the incident flux density from the signal is given by

\begin{equation}
S=\frac{P_{S}}{\Delta f \Omega d^{2}},
\label{eq:S0}
\end{equation}

\noindent
where $P_{S}$ is given by Equation \ref{eq:P_S}, $\Delta f$ is the bandwidth of the signal, usually taken to be half of the \textit{cyclotron frequency}, $f_{C}$ (see Section \ref{sec:fC}), and $\Omega$ is the solid angle through which the signal is emitted.
Radio waves from electromotive moons are emitted through a hollow cone-shaped profile.
In the case of Io, the cone has a wall thickness of $\approx1.5\degree$, and a half angle ranging from 60{\degree} to 90\degree, giving a solid angle of $\approx 0.14-0.16$ steradians \citep{lop07, que01}.
We will assume the upper value of 0.16 steradians to be the general case for exomoons, this will ensure that large flux values will not be artificially favored.

Applying Equations \ref{eq:P_S} and \ref{eq:S0} to Io we obtain a radio power of $\approx4.9$ GW, and a flux density of $\approx2.7$ mJy at 1 light-year from Earth.
Since our efficiency coefficient $\beta_{S}$ is set to $1\%$, then the total dissipated power is $\approx 5\times10^{11}$ W, which is on the same order as values reported in the literature (e.g. see \cite{neu80}, \cite{cra97}, and \cite{zar98}).
The flux density for the other scenarios discussed later can be computed using

\begin{equation}
S=\left(2.7\mbox{ mJy}\right)\left(\frac{P_{S}}{P_{Io}}\right)\left(\frac{24\mbox{ MHz}}{f_{C,max}}\right)\left(\frac{1\mbox{ ly}}{d}\right)^{2},
\label{eq:SS}
\end{equation}

\noindent
where the ratio $P_{S}/P_{Io}$ can be obtained from Figure \ref{fig:P_S}, and $f_{C,max}$ can be calculated from Equation \ref{eq:fc3}.

\section{Detection through Plasma Torus Sharing}\label{sec:share}
In our previous study \citep{noy14}, we focused on how one electromotive moon can be detected if it produces its own plasma.
Now, we will explore how a neighboring exomoon can share the same plasma, and induce radio emissions of detectable amplitude.
This sharing of plasma is possible because the rotation of the planetary magnetic field can drive the co-rotating plasma radially outwards.
Said outwards movement is a result of orbital mechanics: if there is an object in a circular orbit, and then an angular acceleration is applied to said object, it results in a radial acceleration either outwards or inwards, depending on the direction of the angular acceleration.
In our case, the plasma starts from Io's orbit (which is nearly circular) and is accelerated to the co-rotation speed, obtaining an outwards radial velocity in the process.
Furthermore, $V_{co} > V_{orb}$ for all orbits larger than the synchronous orbit (located at $\approx 2.29 R_{J}$), so the Jovian magnetic field continuously accelerates Io's plasma outwards, giving the plasma torus its current shape.
Figure \ref{fig:IPT} shows the Io plasma torus density distribution, replotted from \cite{bag94}\footnote{The data was fitted using a natural cubic spline algorithm; high precision was kept to ensure numerical agreement with the original model.}.
It can be seen from the figure that Io's plasma torus extend a significant distance behind Io, and it is especially dense for distances below 7.5 $R_{J}$.
The sharp decrease in plasma density seen between 7.5 and 8 $R_{J}$, known as the \textit{ramp}, is believed to be caused by Jupiter's ring current \citep{sis81}.

The ionic composition and detailed features of extra-solar plasma tori may vary for each system, but the characteristics that give rise to the general shape of the plasma distribution are quite general.
These characteristics include:
(i) a central mass to orbit, like Jupiter;
(ii) a source of plasma strong enough to sustain a torus, like Io;
(iii) a satellite to interact with the plasma, also like Io; and
(iv) a rotating magnetic field, like on Jupiter.
All these properties are required to obtain a co-rotating, radially-expanding plasma torus whose behavior should be similar to that of Io's.
We will now use this plasma torus to demonstrate that sharing plasma can effectively lead to detection of multiple exomoons.

To illustrate the mechanism, we will substitute Jupiter's moon, Europa, for a hypothetical exomoon.
We will change the exomoon's radius, and orbital distance to simulate the scenarios that could be encountered in exoplanetary systems.
For $R_{S}$ we will use the three values corresponding to Io (1821.3 km), Europa (1560.8 km), and Ganymede (2631 km).
As for the orbital radius, $r_{S}$, the exomoon cannot be too far away because the plasma torus would be too diluted to cause any significant effect, and it cannot be too close to Io ($r_{S}\approx 6.03 R_{J}$) or the system might become orbitally unstable.  For the upper limit, we can simply chose $r_{S}\leq 10 R_{J}$.
For the lower limit, however, more care must be taken to establish criteria which agrees with known stability models.
First, the moon system described here consists of a very large central mass and two orbiting masses of comparable size, so stability criteria used with the restricted three-body problem cannot be applied here.
Perhaps the closest stability results that can be applied to our system are those found for Kepler-36, which 
consists of a Sun-like star, a super-Earth, and a mini-Neptune.  The two planets orbit at 0.1153 and 0.1283 au (respectively), making them the planets with the closest orbits with respect to each other known to date.   

Since the notion of Hill stability has been extended to the general three-body problem (see \cite{mb82}, also
\cite{sm08} and references therein),  the radius of the Hill sphere, $R_{H}$, is given by

\begin{equation}
R_{H}\approx a(1-e)\left(\frac{M_{S}}{3M_{P}}\right)^\frac{1}{3},
\label{eq:hill}
\end{equation}

\noindent
where $a$, $e$, and $M_{S}$ are the semi-major axis, eccentricity, and mass of the moon, respectively.
When Equation \ref{eq:hill} is applied to Io, Europa, and Ganymede the largest Hill radius found is that 
of Ganymede at 0.453 $R_{J}$.  In other words, we can choose our exomoon's smallest orbit to be $6.5 R_{J}$, 
and still have all satellites be outside each other's Hill sphere.  Thus, the range $7 R_{J}\leq r_{S}\leq 
10 R_{J}$ fulfills this last requirement also.  Callisto was excluded since its orbit ($a\approx26.9R_{J}$) 
is too far away from Io to be relevant for this study.  For clarification, this paper does not intend be 
a complete exomoon orbital stability analysis, but only to provide a range of orbital distances where the 
exomoon can reasonably be expected to survive in the long term.

Using Equation \ref{eq:P_S}, we calculated the expected output power for the previously mentioned values of $R_{S}$ for the prograde and retrograde cases.
The results, presented in Figure \ref{fig:P_S}, show that plasma torus sharing can be an extremely important mechanism for making electromotive moons detectable.
For example, if Io is detectable then either a Ganymede-size prograde exomoon, or an Io-size retrograde exomoon would also be detectable up to $r_{S}\approx 7.3R_{J}$ away from Jupiter.
More strikingly, a Ganymede-size retrograde exomoon would be twice as powerful as Io at this distance, and be detectable up to $0.5R_{J}$ farther than the exomoons in the previous example. In other words, this Ganymede-size retrograde exomoon could be detectable even if the plasma source is not.

\section{Signal Frequency, Bandwidth, and Periodicity}\label{sec:fC}
As mentioned before, exomoon radio signals are produced through an electron cyclotron maser process.
The frequency of these cyclotron emissions, the so called cyclotron frequency, is controlled by the strength of the magnetic field at emission site.
Explicitly, the cyclotron frequency, $f_{C}$, is given by

\begin{equation}
f_{C}=\frac{e}{2\pi m_{e}}B,
\label{eq:fc}
\end{equation}

\noindent
where $B$ is the magnetic field at the point of emission, $e$ is the electron charge, and $m_{e}$ is the electron mass. Writing the coefficient in Equation \ref{eq:fc} explicitly yields the more convenient form $f_{C}=(2.8\mbox{ MHz/Gauss})B$.

Since each exomoon in the system is connected to a different field line\footnote{Since $R_{S}<<r_{S}$, we take the field line that passes through the center of the exomoon to be the average field line.}, then we need to know $B$ at any point within that \textit{specific} field line.
If the exomoon is near the exoplanet's rotational equator, then the points on the field line are constrained by the relationship $r=r_{S}\sin^{2}\theta$.
Combining this expression with $\sin^{2}\theta + \cos^{2}\theta=1$, we find that $\cos^{2}\theta=1-r/r_{S}$, which we can substitute into Equation \ref{eq:Bmag} to get

\begin{equation}
B(\vec{r})=\frac{\mu_{0}}{4\pi}\frac{m_{P}}{r^{3}}\left(4-3\frac{r}{r_{S}}\right)^{\frac{1}{2}},
\label{eq:Brs}
\end{equation}

Thereby eliminating the $\theta$ dependence from the magnetic field equation.
Within the path of the electrons traveling between an electromotive moon and the exoplanet, the largest magnetic field is found at the exoplanet's ionosphere (at $r\approx R_{P}$).
Therefore, the maximum emitted cyclotron frequency is emitted at the exoplanet's ionosphere as well.
Combining Equation \ref{eq:fc} with Equation \ref{eq:Brs}, and setting $r=R_{P}$ we find that

\begin{equation}
f_{max}=\frac{\mu_{0}}{8\pi^{2}}\frac{e}{m_{e}}\frac{m_{P}}{R_{P}^{3}}\left(4-3\frac{R_{P}}{r_{S}}
\right)^{\frac{1}{2}}\ ,
\label{eq:fc2}
\end{equation}
\noindent
which is the maximum frequency that can be emitted along the field line connected to the exomoon.  Using Equation 
\ref{eq:m} and simplifying, we get the more convenient expression
\begin{equation}
f_{max} = (4.367\times 10^{30} \mbox{ m\superscript{3}/s})\frac{\left( M_{P}/M_{J}\right)^{1.15}}{R_{P}^{3}}
\left( 4-3\frac{R_{P}}{r_{S}}\right)^{\frac{1}{2}}
\label{eq:totfmax}\ .
\end{equation}
\noindent

Equation \ref{eq:fc2} is asymptotic, approaching a maximum value of $(\mu_{0}em_{P})/(4\pi^{2}m_{e}R_{P}^{3})$ as $r_{S}$ becomes very large.
This value provides a maximum cutoff frequency for all the moons in the system.
If the radius of the exoplanet is not known when calculating $f_{max}$, many authors (e.g. \citep[and references therein]{gri07}) choose to assume $R_{P} \approx R_{J}$.
Studies on the radius-mass relationship of exoplanets show that $R_{P} = R_{J}$ is a good approximation for Jovian exoplanets \citep[and references therein]{mor12}, and therefore we will use this approximation as well.
For the case of Jupiter, with a magnetic dipole moment $m=1.56\times10^{27}$ A•m\superscript{2}, and a radius $R_{J}=69,911$ km, Equation \ref{eq:fc2} can be simplified to

\begin{equation}
f_{max}=(12.8\mbox{ MHz})\left(4-3\frac{R_{J}}{r_{S}}\right)^{\frac{1}{2}},
\label{eq:fc3}
\end{equation}

Using Equation \ref{eq:fc3}, we plotted $f_{max}$ and identified the expected bandwidth for $6R_{J}\leq r_{S}\leq10R_{J}$ on Figure \ref{fig:time}.
It is clear from the plot that $f_{max}$ varies very slowly with $r_{S}$ until it reaches its maximum value of 25.6 MHz, so identifying different exomoons using frequency might be difficult.
To find just how difficult, we need to find what is the largest frequency error, $\delta f_{max}$, that we can admit if we require an uncertainty of $\delta r_{R,f}$ or less in our measurements. 
The two uncertainties are related by $\delta r_{S,f}=(\partial r_{S})/(\partial f_{max})\delta f_{max}$, which we can solve for $\delta f_{C}$ to yield

\begin{equation}
\delta f_{max}=\frac{\delta r_{S,f}}{\partial r_{S}/\partial f_{max}}.
\label{eq:dfc}
\end{equation}

\noindent
Implicitly differentiating Equation \ref{eq:fc2} with respect to $r_{S}$, and solving for the derivative we get

\begin{equation}
\frac{\partial r_{S}}{\partial f_{max}}=\frac{16\pi^{2}m_{e}R_{P}^{2}}{3\mu_{0}em_{P}}\sqrt{r_{S}^{3} \left( 4r_{S} - 3R_{P} \right)},
\label{eq:drdf}
\end{equation}

At first Equation \ref{eq:drdf} appears to be independent from $f_{max}$, but the dependency was simply hidden by the implicit differentiation, and can be recovered by removing the radical with Equation \ref{eq:fc2}.
Combining Equations \ref{eq:fc3} and \ref{eq:drdf} yields

\begin{equation}
\delta f_{max}=\frac{\left( 19.2\mbox{ MHz}\right) R_{J}}{\sqrt{r_{S}^{3} \left( 4r_{S} - 3R_{J} \right)}}\delta r_{S,f}.
\label{eq:dfdrJ}
\end{equation}

This equation predicts that $\delta f_{max}$ is approximately proportional to $1/r_{S}^{2}$, so that exomoons closer to their exoplanet requires less accuracy in the measurement of $f_{max}$.
For example, having $\delta r_{S,f}\leq 0.1R_{J}$ would require $\delta f_{max}\leq 10$ kHz for $r_{S}=10 R_{J}$, but $\delta f_{max}\leq 20$ kHz at $7 R_{J}$.
Nonetheless, we expect bandwidths in the tens of MHz, so an uncertainty of tens of kHz represents an accuracy of less than $0.1\%$, which can be challenging.
In fact, due to the noisy nature of the emission mechanism it is most likely not possible to observe such a small difference in frequency. Nevertheless, exoplanets with larger masses may provide a solution to this problem.
For example, using Equation \ref{eq:totfmax} with two exomoons, one at $6.03R_{J}$ and one at $7.2R_{J}$, for a 
$4M_{J}$ exoplanet, we find that the difference between their frequencies is $\approx 1.3$ MHz. Similarly, if the exoplanet's mass is increased to $8M_{J}$, the frequency difference between the exomoons increases to $\approx 3$ MHz. It is unclear how the increase in planetary mass will affect the natural noisiness of the signal, but $1$ MHz is 50 times larger than the $20$ kHz mentioned earlier, so if the signal noise remains similar to that seen on Jupiter, then we can ascertain that it is easier to distinguish between exomoons around larger exoplanets.

Although identification of multiple exomoons can be done with frequency measurements, the best way to distinguish between exomoons is the differences in the periodicity of their signal's features.
Exomoon signals repeat twice per orbital period, showing maximum intensity whenever the exomoon is approximately to the side of the exoplanet, as seen from Earth.
A clear example of this periodicity is Io, whose signal was found to peak whenever the moon was close to either 93\degree or 246\degree  from the Earth-Jupiter line \citep{big64}.
To assess how accurately we need to measure an exomoon's orbital period, $T_{S}$, we will analyze it as we did for frequency.

The relationship between the required accuracy on $r_{S}$, $\delta r_{S,T}$, and the measured uncertainty on $T_{S}$, $\delta T_{S}$, is

\begin{equation}
\delta T_{S}=\frac{\delta r_{S,T}}{\partial r_{S}/\partial T_{S}}.
\label{eq:dTS}
\end{equation}

\noindent
From Newton's laws, we know that the relationship between $r_{S}$ and $T_{S}$ is given by

\begin{equation}
r_{S}^{3}=\frac{GM_{P}}{4\pi^{2}}T_{S}^{2}.
\label{eq:r3T2}
\end{equation}

\noindent
Differentiating as before, and eliminating $T_{S}$ with Equation \ref{eq:r3T2} we get

\begin{equation}
\frac{\partial r_{S}}{\partial T_{S}}=\frac{1}{3\pi}\sqrt{\frac{GM_{P}}{r_{S}}}.
\label{eq:drdT}
\end{equation}

\noindent
Finally, substituting Equation \ref{eq:drdT} into Equation \ref{eq:dTS}, we find that

\begin{equation}
\delta T_{S}=3\pi\delta r_{S,T}\sqrt{\frac{r_{S}}{GM_{P}}}.
\label{eq:dTS2}
\end{equation}

For our hypothetical exomoon, the orbital period ranges from 2-4 days.
Using Equation \ref{eq:dTS2} with our example, we find that a $\delta r_{S,T}\leq 0.1R_{J}$ can be obtained with a $\delta T_{S}$ as large as 1 hour.
In fact, a $\delta r_{S,T}\leq 0.01R_{J}$ can be attained with a $\delta T_{S}$ of 6 minutes.
Comparing these values to the time scales shown in Figure 6a, we see that 6 minutes of accuracy would not be hard to obtain from Io-DAM emissions (panel a), which have time scales of a few hours. 

Although measuring $T_{S}$ leads to more accurate identifications than frequency measurements, measuring $f_{C,max}$ accurately is still critical because, as we shall see in the next section, these measurements allow us to know some important properties of the system without the need for computational signal analysis.

\section{Physical Properties Derived from Multiple Detections}\label{sec:gifts}

If we divide Equation \ref{eq:r3T2} by $(4/3)\pi R_{P}^{3}$, we find that the density of exoplanet, $D_{P}$, can be expressed as

\begin{equation}
D_{P}=\frac{3\pi}{GT_{S}^{2}}\left(\frac{r_S}{R_{P}}\right)^{3},
\label{eq:DP}
\end{equation}

\noindent
but we need to know the ratio $r_{s}/R_{P}$ in advance.
If the signals of at least two exomoons are detected in a system, we can use that information to calculate that value.
Knowing the maximum frequency and orbital period for two exomoons, $(T_{i}, f_{i})$ and $(T_{j}, f_{j})$, we can use Equation \ref{eq:fc2} to get

\begin{equation}
\left(\frac{f_{i}}{f_{j}}\right)^{2}=\left(\frac{r_{j}}{r_{i}}\right)\frac{4r_{i}-3R_{P}}{4r_{j}-3R_{P}},
\label{eq:f1f2}
\end{equation}

\noindent
which we can further rearrange to obtain

\begin{equation}
\frac{r_{i}}{R_{P}}=\frac{3}{4}\left(\frac{f_{j}^{2}-(r_{i}/r_{j})f_{i}^{2}}{f_{j}^{2}-f_{i}^{2}}\right).
\label{eq:riRP}
\end{equation}

Notice the ratio $r_{i}/r_{j}$ is still needed; however, this ratio can easily be obtained because both exomoons orbit the same exoplanet.
Using Equation \ref{eq:r3T2} with $r_{i}$ and $r_{j}$, we find that

\begin{equation}
\frac{r_{i}}{r_{j}}=\left(\frac{T_{i}}{T_{j}}\right)^{\frac{2}{3}},
\label{eq:rirj}
\end{equation}
\noindent
which provides the last piece of data needed to calculate the $D_{P}$.
Additionally, Equation \ref{eq:riRP} can be solved for either exomoon, and each solution can be used to calculate $D_{P}$ separately.
Hence, we can average the values of $D_{P}$ and further reduce the uncertainty by a factor of $\sqrt{2}$.
Even better, now that we know $r_{s}/R_{P}$, and if and if $R_{P}$ is also known (e.g. from planetary transits), we can use Equation \ref{eq:fc2} to calculate $r_{s}$, $M_{P}$ and $m_{P}$ as well.
In other words, a multiple-exomoon detection allows us to readily calculate almost all the system's physical quantities without the need for complex numerical methods.
One might argue that the previously mentioned noisiness of the signals will make it difficult to measure frequency differences accurately, but it should still give us a rough idea of the configuration of the system. 
In this regard, the situation can be likened to other known cases where planetary properties have not been measured to high accuracy (e.g. OGLE-2006-BLG-109L c has a semi-major axis of $4.5\pm 2.2$ AU \citep{ben10}), and yet still give great insight into the inner structure of the system.

\section{More Plasma Sources}\label{sec:misc}
\begin{figure}
\epsscale{0.80}
\includegraphics[width=1\linewidth]{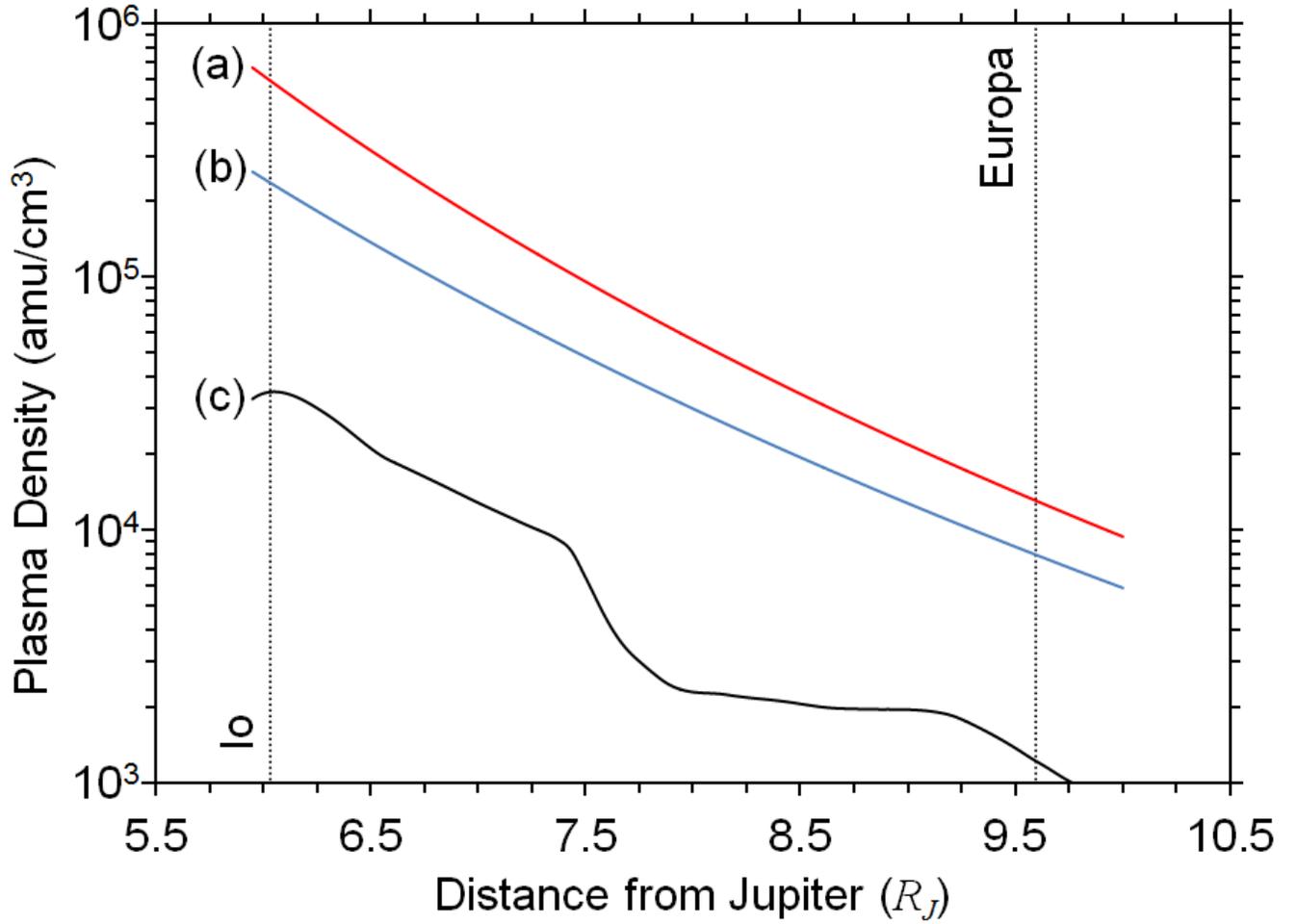}\hfill
\caption{Io's plasma torus density profile, and calculated critical plasma density (see Section \ref{sec:misc}) shown as a function of distance from Jupiter.
Curves (a) and (b) show $\rho_{C}$ calculated for the prograde and retrograde cases, respectively.
Curve (c) shows the total plasma density found between 6 and 10 $R_{J}$[Adapted from \cite{bag94}].
The orbits of Io and Europa are shown for reference. Note: 1 amu/cm\superscript{3} = $1.66\times10^{-21}$ kg/m\superscript{3}, and 1 R\subscript{J} = $69,911$ km.}
\label{fig:IPT}
\end{figure}

\begin{figure}
\epsscale{0.80}
\includegraphics[width=1\linewidth]{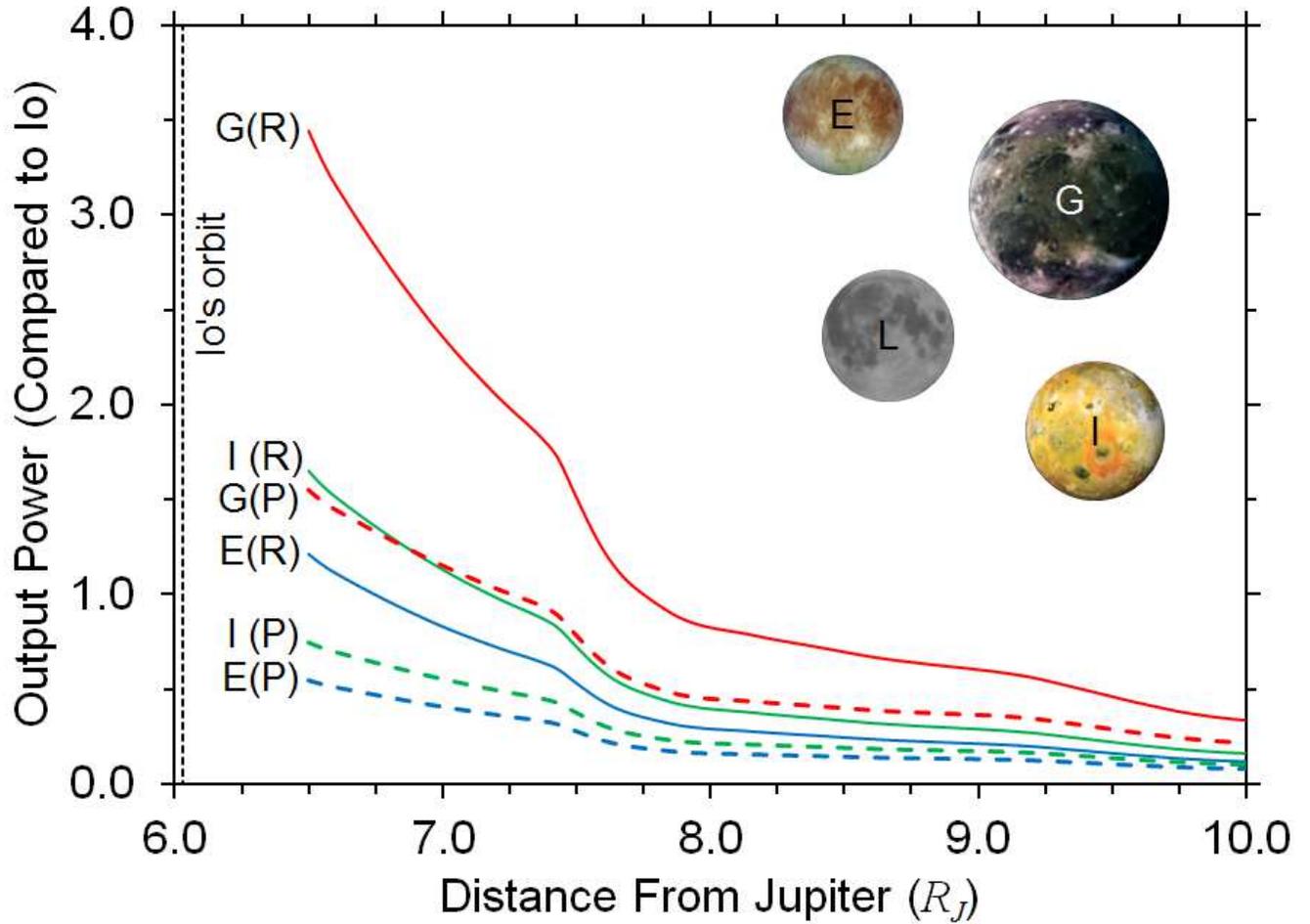}\hfill
\caption{Radio emission output power, $P_{S}$, calculated for various exomoon sizes in orbits between 6.5 and 10 $R_{J}$.
The labels (P) and (R) denote prograde and retrograde configurations.
We used the sizes of the Galilean satellites Europa, Io, and Ganymede as examples (labels E, I, and G respectively).
All the moons used, as well as Earth’s moon, L, are shown to scale for comparison.
Power is given in Io units, where 1 unit $\approx 4.9$ GW, as calculated by our model.}
\label{fig:P_S}
\end{figure}

\begin{figure}
\epsscale{0.80}
\includegraphics[width=1\linewidth]{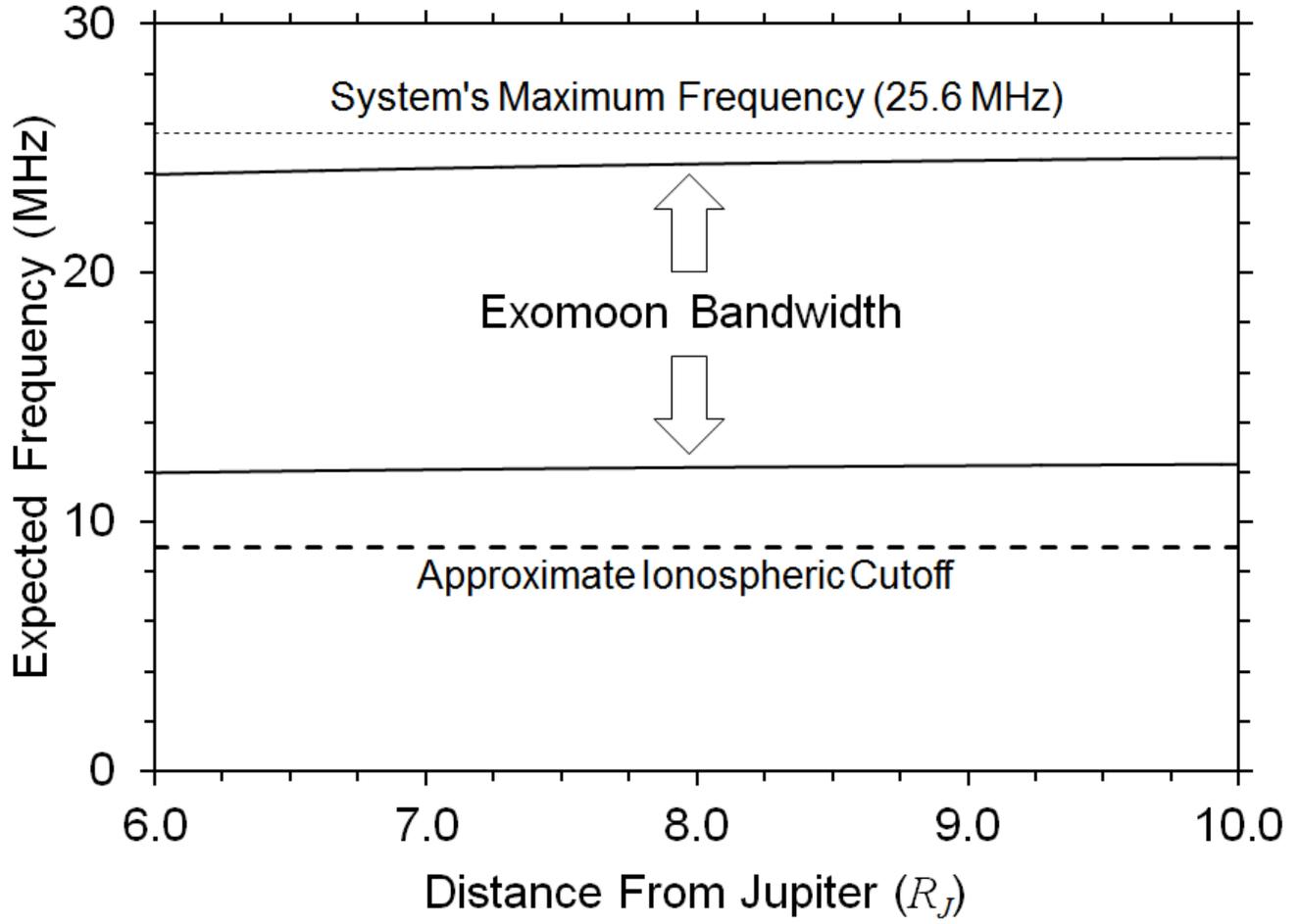}\hfill
\caption{Expected exomoon signal bandwidth versus orbital radius.
The exomoon bandwidths include all frequencies between $f_{C}$ and $f_{C}/2$, with (i) being the
approximate ionospheric cutoff, and (ii) being the maximum possible frequency with a 1 M$_J$ exoplanet
(25.6 MHz).  The maximum possible frequency with a 4 M$_J$ exoplanet is about 126.1 MHz (not shown). 
}
\label{fig:time}
\end{figure}

\begin{figure}
\epsscale{0.80}
\includegraphics[width=1\linewidth]{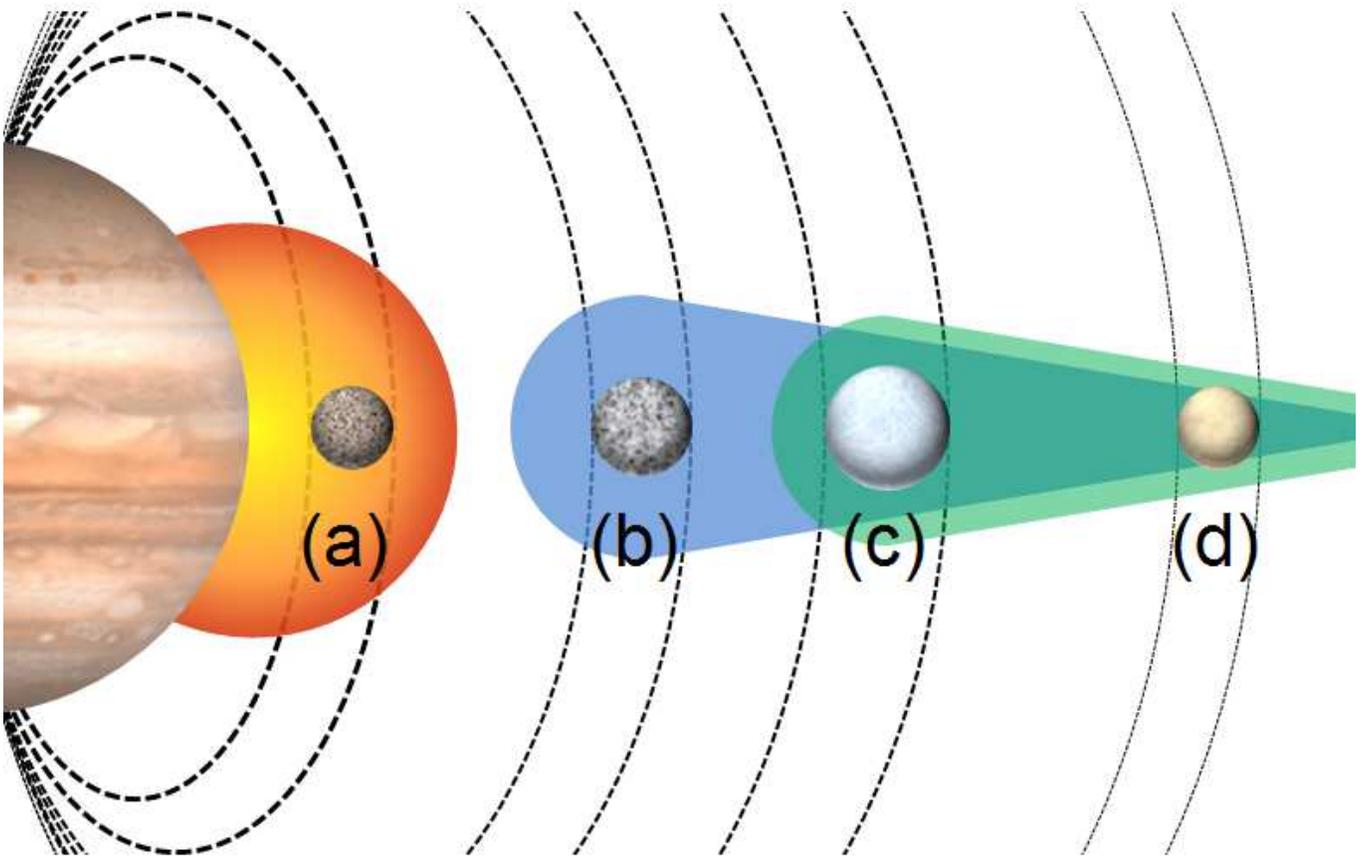}\hfill
\caption{Four pathways through which an exomoon could become a detectable electromotive moon: (a) An exomoon using a hot Jupiter's magnetospheric plasma; (b) An exomoon which, like Io, provides enough plasma to create a dense torus; (c) An exomoon whose plasma torus is a combination of its own ionospheric plasma, and the plasma from a donor exomoon; and (d) An exomoon which relies solely on the plasma from one or more donor exomoons. The dashed curves are the field lines which connect the exomoons to the exoplanet's magnetic poles. \textit{Image Credit:} Jupiter Photo NASA/JPL/Caltech (NASA photo \# PIA00343)}
\label{fig:summary}
\end{figure}

Our previous discussion was focused on one exomoon (the receptor) using plasma from a neighboring exomoon (the donor) to become detectable.
However, the receptor exomoon itself can also be a source of plasma.
For example, \cite{bag94} noted that the increase in oxygen ions seen past 7.5 $R_{J}$ in Figure \ref{fig:IPT} could mean that Europa is also a source of plasma for the system.
Later studies corroborate this statement, and further state that Europa can sustain an average plasma density of $\approx 2500$ amu/cm\superscript{3}, which is much less than Io ($\approx42300$ amu/cm\superscript{3}), but significantly larger than Ganymede ($\approx 54$ amu/cm\superscript{3}), and Callisto ($\approx 1.6$ amu/cm\superscript{3}) \citep{kiv04}.
To illustrate,  the exomoon $c$ on Figure \ref{fig:summary} receives plasma from exomoon $b$ but it also produces its own plasma torus, and uses the combination of both plasmas to generate currents.
In contrast, exomoon $d$ does not produce its own plasma torus, but it can still use the plasma provided by exomoons $b$ and $c$.
Similarly, the exoplanet itself could be the source of plasma for an exomoon that cannot produce its own.
An exomoon of this type, which depends solely on the plasma trapped in the exoplanet’s magnetosphere to produce currents, is illustrated by exomoon $a$ on Figure \ref{fig:summary}.

\begin{figure}
\epsscale{0.80}
\includegraphics[width=1\linewidth]{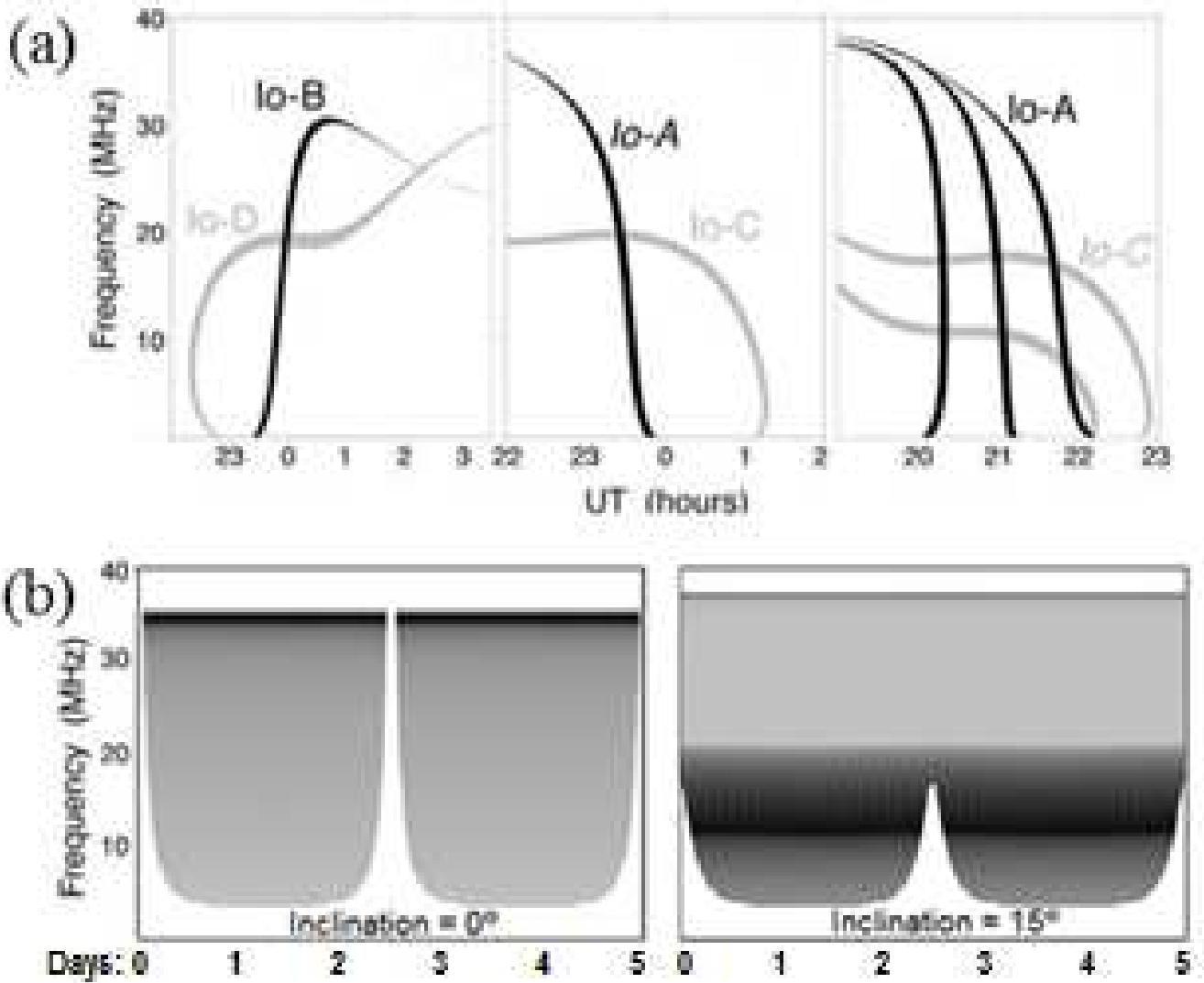}\hfill
\caption{Expected exomoon signal bandwidth versus orbital radius.
Comparison between exomoon and exoplanet signal signatures.
(a) Simulated dynamic spectra of typical Io-Jupiter arc emissions known as Io-A, Io-B, Io-C, and Io-D. Black is Northern Hemisphere and Gray is Southern Hemisphere. (Taken from \cite{hes08}).
(b)Simulated dynamic spectra from an artificial Hot Jovian exoplanet with two orbital inclinations (Taken from \cite{hes11}). Shade darkness represents relative intensity.}
\label{fig:compare}
\end{figure}

The amount of planetary plasma increases with decreasing exoplanet semi-major axis due to increasing stellar XUV irradiation \citep{kos10}.
Thus, stellar irradiation could conceivably make it possible for electromotive moons to be detected in the magnetospheric plasma of hot Jupiters.
Auroral radio emissions equivalent to non-Io-DAM will increase as well \citep[ and references therein]{nic11, zar07}, but given the differences in spectral signatures between Io-DAM and non-Io-DAM we do not expect these emissions to be a significant problem.
Explicitly, exomoon signals should show long, thin arcs lasting several hours, like those on the simulated Io-DAM dynamic spectra shown on Figure \ref{fig:compare}a \citep[Taken from][]{hes08}, whereas exoplanet signals should show a wide-band, near-constant signal like the one shown for a hypothetical hot Jupiter on Figure \ref{fig:compare}b \citep[Taken from][]{hes11}.

In general, we can say that an electromotive moon can obtain plasma from three different sources: (1) its own ionosphere, (2) the plasma torus of another moon, and (3) and its exoplanet's magnetospheric plasma.
To take into account the plasma from all sources, we set the plasma density near an exomoon to be the sum of the plasma from all sources,

\begin{equation}
\rho_{S}(r_{S}) = \sum\limits_{\mbox{All sources}}\rho_{i}(r_{S})
\label{eq:totrho}
\end{equation}

\noindent
and then use that plasma density to calculate the emissions' power and expected flux density.
It is important to note, however, that Equation \ref{eq:P_S} shows asymptotic behavior in its dependance to $\rho_{S}$, and it can reach a saturation point at which the increase in $P_{S}$ with increasing plasma density is negligible.
In other words, if we let $\rho_{S}$ grow to infinity, then $P_{S}$ becomes independent of the plasma density.
The point at which the effects of changing $\rho_{S}$ begin to have little effect on output power is given by

\begin{equation}
\rho_{C}=\frac{1}{\mu_{0}}\left(\frac{B_{S}}{V_{\rho}}\right)^{2}
\label{eq:rhoC}
\end{equation}

\noindent
where we have defined $\rho_{C}$ to be this particular value of $\rho_{S}$. We call $\rho_{C}$ the \textit{critical plasma density}.

Thus, substituting Equation \ref{eq:rhoC} into Equation \ref{eq:P_S}, we can see that when $\rho_{S}$ reaches $\rho_{C}$, $P_{S}$ is already past $70\%$ of its maximum.
Curves (a) and (b) on Figure \ref{fig:IPT} show $\rho_{C}$ for the prograde and retrograde case of our example, respectively. 
From the graph we see that the value of $\rho_{C}$ is at least 1 order of magnitude higher than $\rho_{S}$ at every point on the Io plasma torus, which hints that $\rho_{C}$ can be expected to be generally high.
In other words, it can be safely claimed that more plasma means more radio power from exomoons.

\section{Conclusions}\label{sec:bye}
Our previously proposed idea of single moon detection due to its radio emission caused by the interaction with a Jupiter-like exoplanet \citep{noy14} is now extended to a second exomoon as well as to multiple-exomoon systems.
We suggested that plasma sources other than an exomoon's own ionosphere, such as the plasma torus sharing described here, can lead to the production of radio signals, and the resulting emissions from this process can be used to detect multiple exomoons.
We referred to such exomoons as electromotive moons, and pointed out that they could be common.  

Plasma torus sharing could be the dominant path by which multiple electromotive moons from a single system can become detectable.
The signal from each exomoon in a multiple-exomoon system can be distinguished from the others because of its unique periodicity, and to some extend from its maximum cyclotron frequency. The information obtained from multiple-exomoon detections can be used to infer various physical properties of the system, including each exomoon's orbital radius, and the exoplanet's mass, radius, and magnetic dipole moment.

We thank the anonymous referee for many useful comments and suggestions that helped us to improve our original manucsript.  We would also like to thank Marialis Rosario-Franco for discussions and comments on the manuscript.
Our research on Alfv\'en waves in the exoplanet-exomoon environment was partially supported by NSF under the grant AGS $1246074$ (Z.E.M. and J.P.N.), and GAANN Fellowship (J.P.N.).  Z.E.M. also acknowledges the support of this work by the Alexander von Humboldt Foundation.

{}

\begin{thebibliography}{0}%
\makeatletter
\providecommand \@ifxundefined [1]{%
 \@ifx{#1\undefined}
}%
\providecommand \@ifnum [1]{%
 \ifnum #1\expandafter \@firstoftwo
 \else \expandafter \@secondoftwo
 \fi
}%
\providecommand \@ifx [1]{%
 \ifx #1\expandafter \@firstoftwo
 \else \expandafter \@secondoftwo
 \fi
}%
\providecommand \natexlab [1]{#1}%
\providecommand \enquote  [1]{``#1''}%
\providecommand \bibnamefont  [1]{#1}%
\providecommand \bibfnamefont [1]{#1}%
\providecommand \citenamefont [1]{#1}%
\providecommand \href@noop [0]{\@secondoftwo}%
\providecommand \href [0]{\begingroup \@sanitize@url \@href}%
\providecommand \@href[1]{\@@startlink{#1}\@@href}%
\providecommand \@@href[1]{\endgroup#1\@@endlink}%
\providecommand \@sanitize@url [0]{\catcode `\\12\catcode `\$12\catcode
  `\&12\catcode `\#12\catcode `\^12\catcode `\_12\catcode `\%12\relax}%
\providecommand \@@startlink[1]{}%
\providecommand \@@endlink[0]{}%
\providecommand \url  [0]{\begingroup\@sanitize@url \@url }%
\providecommand \@url [1]{\endgroup\@href {#1}{\urlprefix }}%
\providecommand \urlprefix  [0]{URL }%
\providecommand \Eprint [0]{\href }%
\providecommand \doibase [0]{http://dx.doi.org/}%
\providecommand \selectlanguage [0]{\@gobble}%
\providecommand \bibinfo  [0]{\@secondoftwo}%
\providecommand \bibfield  [0]{\@secondoftwo}%
\providecommand \translation [1]{[#1]}%
\providecommand \BibitemOpen [0]{}%
\providecommand \bibitemStop [0]{}%
\providecommand \bibitemNoStop [0]{.\EOS\space}%
\providecommand \EOS [0]{\spacefactor3000\relax}%
\providecommand \BibitemShut  [1]{\csname bibitem#1\endcsname}%
\let\auto@bib@innerbib\@empty
\end{thebibliography}%


\begin{thebibliography}{}

\bibitem[Acuna et al. (1981)]{acu81} 
Acuna, M. H., Neubauer, F. H., Ness, N. F. 1981, \jgr, 86, 8513

\bibitem[Agnor \& Hamilton (2006)]{agn06}
Agnor, C. B., Hamilton, D. P., 2006, \nat, 441, 7090, 192-194

\bibitem[Bagenal (1994)]{bag94}
Bagenal, F., 1994, \jgr, 99, A6

\bibitem[Bagenal \& Delamere (2011)]{bag11}
Bagenal, F., Delamere, P. A., 2011, \jgr, 116, A05209

\bibitem[Belcher (1987)]{bel87} 
Belcher, J.  1987, Science, 238, 170

\bibitem[Ben-Jaffel et al. (2014)]{benj14}
Ben-Jaffel, L., Ballester, G. E., 2014, \apj, 785, L30

\bibitem[Bennett et al. (2014)]{ben14}
Bennett, D. P., Batista, V., Bond, I. A., \& 47 other authors, 2014, \apj, 785, 15

\bibitem[Bennett et al. (2010)]{ben10}
Bennett, D. P., Rhie, S. H., Nikolaev, S., \& 71 other authors, 2010, \apj, 713, 837

\bibitem[Bigg (1964)]{big64} 
Bigg, E. K.,  1964, \nat, 203, 1008

\bibitem[Bose et al. (2008)]{bos08}
Bose, S. K., Sarkar, S., Bhattacharyya, A. B., 2008, Indian Journal of Radio \& Space Physics, 37, 77

\bibitem[Canup \& Ward. (2006)]{can06}
Canup, R. M., Ward, W. R., 2006, \nat 441

\bibitem[Crary (1997)]{cra97} 
Crary, F. J.,  1997, \jgr, 102, 37

\bibitem[Durand-Manterola (2009)]{dur09} 
Durand-Manterola, H. J.,  2009, Planet. Space Sci., 57, 1405

\bibitem[Farrell et al. (1999)]{far99}
Farrell, W. M., Desch, M. D., Zarka, P., 1999, \jgr, 104, E6, 14025-14032

\bibitem[Goldreich et al. (1969)]{gol69}
Goldreich, P., Lynden-Bell, D., 1969, \apj, 156, 59

\bibitem[Griessmeier et al. (2007)]{gri07}
{Grie{\ss}meier}, J.-M. and {Zarka}, P. and {Spreeuw}, H. 2007, \aap, 475, 359

\bibitem[Heller et al. (2014)]{hel14}
Heller, R., Williams, D., Kipping, D., Limbach, M. A., Turner, E., Greenberg, R., Sasaki, T., Bolmont, É., Grasset, O., Lewis, K., Barnes, R., Zuluaga, J. I., 2014, Astrobiology, Vol. 14, Issue 9, p. 798-835

\bibitem[Heller \& Pudritz (2015)]{hel15}
Heller, R., Pudritz, R., 2015, \aap 578, A19

\bibitem[Hess et al. (2008)]{hes08}
Hess, S.,  Cecconi, B., Zarka, P., 2008, Geophysical Research Letters, 35, L13107

\bibitem[Hess \& Zarka (2011)]{hes11}
Hess, S. L. G., Zarka, P., 2011, \aap, 531, A29

\bibitem[Hughes (2003)]{hug03}
Hughes, D. W., 2003, Planet. Space Sci., 51, 517

\bibitem[Johnson \& Huggins (2006)]{joh06}
Johnson, R. E., Huggins, P. J., 2006, Astronomical Society of the Pacific, 118, 846, 1136-1143

\bibitem[Kipping et al. (2009)]{kip09}
Kipping, D. M., Fossey, S. J., Campanella, G. 2009, \mnras, 400, 398

\bibitem[Kivelson et al. (2004)]{kiv04} 
Kivelson, M. G.,  2004, Jupiter (Cambridge, Cambridge University Press), Chapter 21

\bibitem[Koskinen et al. (2010)]{kos10}
Koskinen, T. T., Yelle, R. V., Lavvas, P., Lewis, N. K., 2010, \apj, 723, 116

\bibitem[Lopes \& Spencer (2007)]{lop07} 
Lopes, R. M., \& Spencer, J. R.  2007, Io After Galileo: A New View of Jupiter's Volcanic Moon (Geophysical Sciences, Springer Praxis Books) 

\bibitem[Marchal \& Bozis (1982)]{mb82}
Marchal, C., \& Bozis, G. 1982, Celest. Mech., 26, 311 

\bibitem[Mauk et al. (2001)]{mau01} 
Mauk, B. H., Williams, D. J., Eviatar, A. 2001, \jgr, 106, 26195

\bibitem[Mordasini et al. (2012)]{mor12}
Mordasini, C., Alibert, Y., Georgy, C., Dittkrist, K. M., Klahr, H., and Henning, T., 2012, A\&A, 547, A112

\bibitem[Neubauer (1980)]{neu80} 
Neubauer, F. M.  1980, \jgr, 85, 1171

\bibitem[Nichols (2011)]{nic11}
Nichols J. D. 07/2011, \mnras, 414, 2125

\bibitem[Noyola et al. (2014)]{noy14}
Noyola, J. P., Satyal, S., Musielak, Z. E., 2014, \apj, 791, 25

\bibitem[Porter \& Grundy (2011)]{por11}
Porter, S. B., Grundy, W. M., 2011, \apj 736 L14

\bibitem[Queinnec et al. (2001)]{que01}
Queinnec, J., Zarka, P. 2001, Planet. Space Sci., 49, 365-376

\bibitem[Saur et al. (1999)]{sau99} 
Saur, J., Neubauer, F. M., Strobel, D. F., Summers, M. E., 1999, \jgr, 104, 25105

\bibitem[Siscoe et al (1981)]{sis81}
Siscoe, G. L., Eviatar, A., Thorne, R. M., Richardson, J. D., Bagenal, F., Sullivan, J. D., 1981, \jgr, 86, 8480

\bibitem[Snellen et al. (2014)]{sne14} 
Snellen, I. A. G., Brandl, B. R., de Kok, R. J., Brogi, M., Birkby, J., Schwarz, H., 2014, \nat, 509, 63

\bibitem[Su (2009)]{su09} 
Su, Y. -J.,  2009, IAU Symp., Eds K. G. Strassmeier, A. G. Kosovichev, and J. E. Beckman, Vol. 259, p. 271

\bibitem[Szenkovits \& Mak\'o (2008)]{sm08}
Szenkovits, F., \& Mak\'o, Z. 2008, Celest. Mech. Dyn. Astron., 101, 273 
 
\bibitem[Zarka (1998)]{zar98} 
Zarka, P., 1998, \jgr, 103, 20159

\bibitem[Zarka et al. (2001)]{zar01} 
Zarka, P., Treumann, R. A., Ryabov, B. P., Ryabov, V. B., 2001, Astrophys. Space Sci., 277, 293

\bibitem[Zarka et al. (2004)]{zar04} 
Zarka, P., Cecconi, B., Kurth, W. S., 2004, \jgr: Space Physics, 109, A9, A09S15

\bibitem[Zarka (2007)]{zar07} 
Zarka, P., 2007, Planet. Space Sci., 55, 598

\end{thebibliography}
\end{document}